\documentclass[fleqn,usenatbib]{mnras}
\usepackage{newtxtext,newtxmath}

\usepackage[T1]{fontenc}
\usepackage{ae,aecompl}

\usepackage{graphicx}	
\usepackage{amsmath}	
\usepackage{amssymb}	

\newcommand{\ms}{\mbox{m\,s$^{-1}$}}

\newcommand{\Mjup}{\mbox{$M_{\rm Jup}$}}

\newcommand{\REarth}{\mbox{$R_{\oplus}$}}

\newcommand{\ltsimeq}{\raisebox{-0.6ex}{$\,\stackrel
         {\raisebox{-.2ex}{$\textstyle <$}}{\sim}\,$}}

\begin{document}


\title[Cool Jupiters]{Cool Jupiters greatly outnumber their toasty siblings: Occurrence rates from the Anglo-Australian Planet Search}

\author[R.A. Wittenmyer et al.]{
Robert A. Wittenmyer,$^{1}$\thanks{E-mail: rob.w@usq.edu.au}
Songhu Wang$^{2,3}$
Jonathan Horner$^{1}$
R.P. Butler$^{4}$
\newauthor C.G. Tinney$^{5}$
B.D. Carter$^{1}$
D.J. Wright$^{1}$
H.R.A. Jones$^{6}$
J. Bailey$^{5}$
S.J. O'Toole$^{7,8}$
\newauthor Daniel Johns$^{9}$
\\
$^{1}$University of Southern Queensland, Centre for Astrophysics, USQ Toowoomba, QLD 4350 Australia\\
$^{2}$Department of Astronomy, Yale University, 52 Hillhouse Avenue, New Haven, CT 06511, USA \\
$^{3}$51 Pegasi Fellow \\
$^{4}$Department of Terrestrial Magnetism, Carnegie 
Institution of Washington, 5241 Broad Branch Road, NW, Washington, DC 
20015-1305, USA \\
$^{5}$School of Physics and Australian Centre for 
Astrobiology, University of New South Wales, Sydney 2052, Australia \\
$^{6}$Centre for Astrophysics Research, University of 
Hertfordshire, College Lane, Hatfield, Herts AL10 9AB, UK \\
$^{7}$Australian Astronomical Observatory, 105 Delhi Rd, North Ryde, NSW 2113, Australia \\
$^{8}$Australian Astronomical Optics, Faculty of Science and Engineering, Macquarie University, 105 Delhi Rd, North Ryde, NSW 2113, Australia \\
$^{9}$Department of Physical Sciences, Kutztown University, Kutztown, PA 19530, USA
}

\date{Accepted XXX. Received YYY; in original form ZZZ}

\pubyear{2019}

\label{firstpage}
\pagerange{\pageref{firstpage}--\pageref{lastpage}}
\maketitle

\begin{abstract}
Our understanding of planetary systems different to our own has grown dramatically in the past 30 years.  However, our efforts to ascertain the degree to which the Solar system is abnormal or unique have been hindered by the observational biases inherent to the methods that have yielded the greatest exoplanet hauls.  On the basis of such surveys, one might consider our planetary system highly unusual - but the reality is that we are only now beginning to uncover the true picture.  In this work, we use the full eighteen-year archive of data from the Anglo-Australian Planet Search to examine the abundance of 'Cool Jupiters' - analogs to the Solar system's giant planets, Jupiter and Saturn.  We find that such planets are intrinsically far more common through the cosmos than their siblings, the hot Jupiters.  We find that the occurrence rate of such 'Cool Jupiters' is $6.73^{+2.09}_{-1.13}$\%, almost an order of magnitude higher than the occurrence of hot Jupiters (at $0.84^{+0.70}_{-0.20}$\%).  We also find that the occurrence rate of giant planets is essentially constant beyond orbital distances of $\sim$1\,au.  Our results reinforce the importance of legacy radial velocity surveys for the understanding of the Solar system's place in the cosmos. 

\end{abstract}




\section{Introduction}

The story of the Exoplanet Era has been one of continual surprise.  With each new discovery technique, and each new observing facility, planets have been discovered that fail to conform to our expectations of what a planetary system should look like.  From the warm and hot Jupiters and pulsar planets that marked our entry to the era \citep[e.g.][]{psr1257,51peg,HotJupiters}, to the highly eccentric worlds found by radial velocity surveys \citep[e.g.][]{ecc0,ecc1, ecc2, 76920}, as more planets have been found, we have discovered that the population is far more diverse than we could have possibly imagined \citep[see e.g. the review by][]{winn15}.

In the one planetary system we can study in depth, we also see great diversity \citep[e.g.][{\textit{submitted}}]{SotA}.  From small, over-dense planets locked in spin-orbit resonance \citep[Mercury, e.g.][]{Cam88,Benz07,Chau18} to giants with vast numbers of satellites \citep[Jupiter and Saturn, e.g.][]{irreg1,irreg2,irreg3}, it is apparent that the complexity of planetary systems only increases as more information is gleaned.

Despite the fact that we are now thirty years into the Exoplanet Era, we are only now beginning to find systems that truly resemble our own \citep[e.g.][]{5au,anal1,30177,ag18,buch18}.  When one considers the distribution of all exoplanets found to date, and compares it with the planets in the Solar system, the only two which we would have had any chance of discovering to date are the gas giants Jupiter and Saturn \citep{SotA}.  With orbital periods of 12 and 29 years, our two giant planets could be considered a pair of ``Cool Jupiters''.  With this motivation, in this work we define ``cool Jupiters'' as planets with masses greater than 0.3\Mjup\ and orbital periods longer than 100 days.  Such planets are definitively gas giants, analogous to our own Jupiter and Saturn, and this mass boundary is compatible with the detection limits achievable by long-term radial velocity (RV) surveys; Saturn imposes an RV signal of $K\sim$3\,\ms\ over its 29-year period.  

As the dynamically dominant and most readily detectable planets in our Solar system, it is natural to ask whether planetary systems with architectures like our own (i.e. with the most massive planets moving on relatively distant, long-period orbits) are common, or are unusual.  Answering this question will provide the first step towards a wider understanding of the uniqueness of the Solar system in the context of the wider exoplanet population, and is tied to an understanding of the frequencies of Jupiter analogues \citep{jupiters2} and that of Earth-like planets.  Given the widely discussed influence of such giant planets on the formation, evolution, and potential habitability of Earth-like planets within the same system \citep[e.g.][]{raymond06,fof1,fof2,HabRev,fof4,fof3,milank,Graz16,raymond17,bryan19}, the answer is clearly of wide interest.

There is, however, a problem.  The vast majority of newly discovered exoplanets move on orbits that huddle close to their host stars, and precious few are found on orbits that take years or decades to complete  \citep[e.g.][]{witt19, rickman19, kane19}.  These discovery statistics are primarily the result of the biases inherent in the predominant discovery technique - the search for transiting worlds \citep[e.g.][]{exoplanethandbook}.  Both {\it Kepler} and {\it TESS}, the most successful planet discovery factories of the past decade and coming years, respectively, are heavily biased towards finding planets moving on orbits with periods measured in days and weeks, rather than years and decades \citep[e.g.][]{kep1,kep2,kep3,tess1,tess2,tess3}.

To detect planets analogous to Jupiter and Saturn, different techniques must come to the fore.  Around young stars, where planets are still self-luminous from the accrued heat of their accretion, such worlds can potentially be found through direct imaging \citep[e.g.][]{HR8799,HR8799b,Fomal,betapic,Sallum, bowler16}. However, that science is still very much in its infancy, and precious few worlds have been discovered to date\footnote{As of 15 October 2019, just 47 of the 4073 known exoplanets were discovered through direct imaging, based on data from the NASA Exoplanet Archive.}.  In much the same vein, long term studies of the astrometric motion of stars should yield a harvest of Jovian planets - with some authors predicting that the {\it Gaia} spacecraft could find tens of thousands of such planets in the coming decade \citep{perryman14}.  Again, however, that science is still in its youth - with no definitively planetary objects having yet been discovered astrometrically.

Fortunately, there exists another solution.  The radial velocity (RV) surveys that launched the Exoplanet Era, three decades past, have now achieved temporal baselines and precisions sufficient to reveal planets akin to those in our own Solar system.  Few such ``legacy'' surveys are still operating, with the Anglo-Australian Planet Search \citep[AAPS; e.g.][]{aaps1,aaps2,30177} having been brought to a close in 2014 for a total time baseline of 18 years.  The McDonald Observatory planet search is one of the last remaining legacy surveys still in regular operation, with more than 20 years of precise RVs from the 2.7m Harlan J. Smith Telescope \citep[e.g.][]{McD1,McD2}.  That program continues to deliver valuable new discoveries of long-period giant planets \citep[e.g.][]{robertson12, 204313, endl16, blunt19}.  The CORALIE survey has also been running for more than 20 years \citep{Coralie, marmier13, rickman19}.  Despite the untimely demise of AAPS and similar programs, the legacy data sets they provide offer the possibility to investigate both the occurrence of cool giants, and the frequency with which they occur in pairs. 

In this paper, we use the complete AAPS data set to explore the occurrence rate and distribution of giant planets at all orbital separations probed by the nearly two decades of available RV data for a statistically significant sample of Solar-type stars.  Our previous efforts \citep{jupiters1,jupiters2} considered simply the true ``Jupiter analogs'': giant planets with $a>3$\,au moving on low-eccentricity orbits like our own Jupiter. In doing so, we hope to provide data that not only addresses the question of how common or unusual are planetary systems like our own, but also helps to inform our understand of the formation and evolution of giant planets in a broader context.


In section two, we detail the observational data available to use, and describe the methods used to determine the occurrence rates of ''cool Jupiters'' from that data.  In section three, we present and discuss our results, before drawing our conclusions in section four.

\section{Observational Data and Simulation Methods}

We consider here a subsample of the AAPS target list which satisfies the following two criteria: an observational baseline longer than eight years, and more than 30 observations. This is the same selection that was used for our previous work on the occurrence rate of Jupiter analogs \citep{jupiters1, jupiters2}, and yields a sample of 203 stars.  

\subsection{Giant Planets in the Sample}

This sample is known to contain 38 giant planets orbiting 30 stars, with 33 of these considered as ``cool Jupiters'', as per our definition above.  Given that the orbital parameters of most of these planets have not been updated since their discovery, up to a decade ago, we wish to first refine the parameters using the full extent of the latest available data.  Table~\ref{tab:rvdata} gives the details of the various RV data sets used in our fitting.  For HD\,114613, HD\,134987, and HD\,159868, we used the corrected Keck/HIRES RVs \citep{butler17} given in \citet{talor19}.  Publicly-available HARPS DRS velocities were obtained from the ESO Archive, accessed on 2019 Feb 1. For all data sets, where there were multiple observations in a single night, we binned them together using the weighted mean value of the velocities in each night.  We adopted the quadrature sum of the rms about the mean and the mean internal uncertainty as the error bar of each binned point.  We then used the \textit{Systemic Console 2.2000} \citep{mes09} to obtain new Keplerian orbit fits for the 33 cool Jupiters in our sample, with uncertainties derived from 10,000 bootstrap iterations.  The refined orbital parameters are given in Table~\ref{tab:planetparams}.  For the HD\,30177, HD\,39091, and HD\,73526 systems, we cite the most recent published solutions as no new data are available at present.

\begin{table}
	\centering
	\caption{Properties of RV data used for refining orbits.}
	\label{tab:rvdata}
	\begin{tabular}{llll} 
		\hline
		Star & Instrument & $N_{epochs}$ & Reference \\
		\hline
		HD 142 & AAT & 93 & This work \\
		       & ESO/HARPS & 12 & ESO Archive \\
		HD 2039 & AAT & 46 & This work \\
		HD 13445 & AAT & 74 & This work \\
		HD 17051 & AAT & 38 & This work \\
		         & ESO/HARPS & 33 & ESO Archive \\
		HD 20782 & AAT & 57 & This work \\
		         & PARAS & 5  & \citet{kane16} \\ 
		         & ESO/HARPS & 50  & ESO Archive \\
		HD 23079 & AAT & 40 & This work \\
		         & ESO/HARPS & 26 & ESO Archive \\
		HD 23127 & AAT & 44 & This work \\
		          & ESO/HARPS & 24 & ESO Archive \\
		HD 27442 & AAT & 104 & This work \\
		HD 30177 & AAT & 43 & This work \\ 
		         & ESO/HARPS & 41 & ESO Archive \\
		HD 38283 & AAT & 67 & This work \\ 
		         & ESO/HARPS & 5 & ESO Archive \\
		HD 39091 & AAT & 77 & This work \\ 
		         & ESO/HARPS &  & \citet{huang18} \\
		HD 70642 & AAT & 51 & This work \\
		         & ESO/HARPS & 29  & ESO Archive \\
		HD 73526 & AAT & 36 & This work \\ 
		HD 75289 & AAT & 50 & This work \\
		         & CORALIE & 58 & \citet{udry00} \\
		HD 83443 & AAT & 25 & This work \\
		         & CORALIE & 215 & \citet{mayor04} \\
		         & Keck/HIRES & 35 & \citet{butler17} \\
		HD 108147 & AAT & 58 & This work \\
		          & CORALIE & 117 & \citet{pepe02} \\
		HD 114613 & AAT & 244 & This work \\
		          & Keck/HIRES & 37 & \citet{butler17} \\
		          & ESO/HARPS & 27  & ESO Archive \\
		HD 134987 & AAT & 77 & This work \\
		          & Keck/HIRES & 94 & \citet{butler17} \\
		          & ESO/HARPS & 14  & ESO Archive \\
		HD 154857 & AAT & 45 & This work \\
		HD 159868 & AAT & 52 & This work \\
		          & Keck/HIRES & 34 & \citet{butler17} \\
		HD 160691 & AAT & 180 & This work \\
		          & ESO/HARPS & 161  & ESO Archive \\
		HD 179949 & AAT & 66 & This work\\
		          & McD 2.7m & 17 & \citet{witt07} \\
		          & Keck/HIRES & 31 & \citet{butler17} \\
		HD 187085 & AAT & 75 & This work \\ 
		HD 196050 & AAT & 57 & This work \\
		          & ESO/HARPS & 38  & ESO Archive \\
		HD 208487 & AAT & 49 & This work \\
		          & ESO/HARPS & 25  & ESO Archive \\
		HD 213240 & AAT & 37 & This work \\
		HD 216435 & AAT & 79 & This work \\
		          & ESO/HARPS & 13  & ESO Archive \\
		HD 216437 & AAT & 58 & This work \\
		          & ESO/HARPS & 30  & ESO Archive \\
		HD 219077 & AAT & 72 & This work \\
		          & ESO/HARPS & 22  & ESO Archive \\
		GJ 832 & AAT & 39 & \citet{gj832} \\ 
		       & ESO/HARPS & 54  & \citet{gj832} \\
		       & Magellan/PFS & 16  & \citet{gj832} \\
		\hline
	\end{tabular}
\end{table}


\begin{table*}
	\centering
	\caption{Updated orbital solutions for cool Jupiters from the AAPS sample.}
	\label{tab:planetparams}
	\begin{tabular}{llllllll} 
		\hline
		       & Period & Eccentricity & $\omega$ & $T_0$ & $K$ & m sin $i$ & $a$ \\
		Planet & days &   & degrees & BJD-2400000 & \ms & \Mjup & au \\
		\hline
		HD 142b & 352.48$\pm$0.77 & 0.294$\pm$0.076 & 303$\pm$25 & 49876$\pm$122 & 33.2$\pm$2.8 & 1.268$\pm$0.107 & 1.0467$\pm$0.0015 \\
		HD 142c & 6268$\pm$32 & 0.138$\pm$0.055 & 277$\pm$27 & 44649$\pm$2249 & 51.8$\pm$2.5 & 5.35$\pm$0.27 & 7.139$\pm$0.025 \\
		HD 2039b & 1110.1$\pm$3.9 & 0.637$\pm$0.011 & 341.7$\pm$4.4 & 49918$\pm$412 & 106$\pm$48 & 4.5$\pm$1.7 & 2.184$\pm$0.006 \\
		HD 13445b & 15.76480$\pm$0.00004 & 0.048$\pm$0.002 & 269.7$\pm$3.3 & 49997$\pm$6 & 619.2$\pm$1.7 & 6.588$\pm$0.018 & 0.114340$\pm$0.000001 \\
		HD 17051b & 308.3$\pm$1.6 & 0.177$\pm$0.080 & 62$\pm$30 & 49873$\pm$112 & 69.4$\pm$4.6 & 2.49$\pm$0.18 & 0.935$\pm$0.003 \\
		HD 20782b & 597.12$\pm$0.07 & 0.952$\pm$0.004 & 144.1$\pm$1.4 & 49683$\pm$211 & 117.0$\pm$6.3 & 1.457$\pm$0.049 & 1.37400$\pm$0.00014 \\ 
		HD 23079b & 724.5$\pm$2.2 & 0.087$\pm$0.031 & 19$\pm$25 & 49893$\pm$270 & 54.2$\pm$1.3 & 2.41$\pm$0.06 & 1.586$\pm$0.003 \\
		HD 23127b & 1219.5$\pm$10.3 & 0.318$\pm$0.067 & 187$\pm$14 & 49970$\pm$454 & 28.5$\pm$1.7 & 1.54$\pm$0.10 & 2.326$\pm$0.013 \\
		HD 27442b & 429.1$\pm$0.7 & 0.057$\pm$0.037 & 231$\pm$45 & 49705$\pm$157 & 32.1$\pm$1.5 & 1.55$\pm$0.07 & 1.269$\pm$0.001 \\
		HD 30177b$^{1}$ & 2524.4$\pm$9.8 & 0.184$\pm$0.012 & 31$\pm$3 & 51434$\pm$29 & 126.3$\pm$1.5 & 8.08$\pm$0.10 & 3.58$\pm$0.01 \\
		HD 30177c & 11613$\pm$1837 & 0.22$\pm$0.14 & 19$\pm$30 & 48973$\pm$1211 & 70.8$\pm$29.5 & 7.6$\pm$3.1 & 9.89$\pm$1.04 \\
		HD 38283b & 361.0$\pm$1.1 & 0.474$\pm$0.136 & 188$\pm$23 & 49842$\pm$132 & 8.9$\pm$1.6 & 0.289$\pm$0.034 & 1.020$\pm$0.002 \\
		HD 39091b$^{2}$ & 2093.07$\pm$1.73 & 0.637$\pm$0.002 & 330.6$\pm$0.3 & 45852.0$\pm$3.0 & 192.6$\pm$1.4 & 10.02$\pm$0.15 & 3.10$\pm$0.02 \\
		HD 70642b & 2148.7$\pm$9.8 & 0.186$\pm$0.051 & 276$\pm$14 & 49750$\pm$784 & 28.0$\pm$1.5 & 1.75$\pm$0.09 & 3.263$\pm$0.010 \\
		HD 73526b$^{3}$ & 189.65$\pm$0.21 & 0.265$\pm$0.021 & 198.3$\pm$3.6 & 51156.8$\pm$2.6 & 85.4$\pm$2.3 & 2.35$\pm$0.12 & 0.65$\pm$0.01 \\
		HD 73526c & 376.93$\pm$0.69 & 0.198$\pm$0.029 & 294.5$\pm$11.3 & 51051.5$\pm$9.4 & 62.3$\pm$1.8 & 2.19$\pm$0.12 & 1.03$\pm$0.02 \\
		HD 75289b & 3.50916$\pm$0.00002 & 0.062$\pm$0.022 & 154$\pm$21 & 49998.7$\pm$1.0 & 54.5$\pm$1.1 & 0.456$\pm$0.010 & 0.047859$\pm$0.000002 \\
		HD 83443b & 2.98566$\pm$0.00005 & 0.03$\pm$0.02 & 236$\pm$43 & 49997.5$\pm$1.1 & 53.8$\pm$1.3 & 0.379$\pm$0.009 & 0.040464$\pm$0.000001 \\
		HD 108147b & 10.9001$\pm$0.0004 & 0.52$\pm$0.05 & 306$\pm$9 & 49992.8$\pm$4.0 & 29.6$\pm$2.4 & 0.31$\pm$0.02 & 0.101429$\pm$0.000002 \\
		HD 114613b & 3969$\pm$204 & 0.42$\pm$0.09 & 208$\pm$16 & 48351$\pm$1476 & 4.67$\pm$0.39 & 0.384$\pm$0.047 & 5.29$\pm$0.18 \\
		HD 134987b & 258.21$\pm$0.03 & 0.231$\pm$0.006 & 356.0$\pm$2.1 & 49864$\pm$81 & 49.2$\pm$0.4 & 1.556$\pm$0.012 & 0.80803$\pm$0.00007 \\
		HD 134987c & 5358$\pm$31 & 0.092$\pm$0.045 & 295$\pm$27 & 47167$\pm$1761 & 8.95$\pm$0.41 & 0.795$\pm$0.036 & 6.100$\pm$0.023 \\
		HD 154857b & 408.59$\pm$0.45 & 0.467$\pm$0.018 & 57.4$\pm$3.0 & 49958$\pm$152 & 48.5$\pm$1.2 & 2.25$\pm$0.06 & 1.2912$\pm$0.0009 \\
		HD 154857c & 3515$\pm$126 & 0.074$\pm$0.054 & 353$\pm$32 & 49103$\pm$1309 & 23.6$\pm$1.1 & 2.53$\pm$0.14 & 5.42$\pm$0.13 \\
		HD 159868b & 1182.5$\pm$7.0 & 0.006$\pm$0.024 & 185$\pm$122 & 49916$\pm$443 & 38.4$\pm$1.3 & 2.12$\pm$0.07 & 2.252$\pm$0.009 \\
		HD 159868c & 351.9$\pm$1.2 & 0.121$\pm$0.045 & 299$\pm$37 & 49916$\pm$443 & 20.8$\pm$1.2 & 0.758$\pm$0.045 & 1.003$\pm$0.002 \\
		HD 160691e & 308.94$\pm$0.48 & 0.102$\pm$0.020 & 207$\pm$68 & 49855$\pm$53 & 13.4$\pm$0.7 & 0.487$\pm$0.025 & 0.936$\pm$0.001 \\
		HD 160691b & 644.68$\pm$0.50 & 0.052$\pm$0.018 & 25$\pm$62 & 49518$\pm$27 & 36.2$\pm$0.3 & 1.684$\pm$0.013 & 1.530$\pm$0.001 \\
		HD 160691c & 4043$\pm$40 & 0.045$\pm$0.015 & 5$\pm$64 & 47361$\pm$484 & 23.9$\pm$0.5 & 2.050$\pm$0.037 & 5.202$\pm$0.030 \\
		HD 179949b & 3.09254$\pm$0.000017 & 0.042$\pm$0.014 & 204$\pm$25 & 49999$\pm$1 & 113.4$\pm$1.9 & 0.908$\pm$0.015 & 0.043923$\pm$0.000001 \\
		HD 187085b & 1039.4$\pm$12.1 & 0.157$\pm$0.083 & 100$\pm$34 & 48987$\pm$389 & 14.4$\pm$1.2 & 0.776$\pm$0.068 & 2.100$\pm$0.016 \\
		HD 196050b & 1400.1$\pm$8.0 & 0.162$\pm$0.009 & 166$\pm$10 & 49038$\pm$522 & 50.2$\pm$0.6 & 2.924$\pm$0.039 & 2.537$\pm$0.010 \\
		HD 208487b & 129.3$\pm$0.1 & 0.30$\pm$0.07 & 88$\pm$15 & 49970$\pm$48 & 17.3$\pm$1.5 & 0.442$\pm$0.037 & 0.5186$\pm$0.0003 \\
		HD 213240b & 872.4$\pm$2.1 & 0.424$\pm$0.011 & 206.9$\pm$2.1 & 49842$\pm$321 & 95.8$\pm$1.4 & 4.468$\pm$0.071 & 1.871$\pm$0.003 \\
		HD 216435b & 1328$\pm$20 & 0.481$\pm$0.076 & 25$\pm$57 & 49815$\pm$494 & 21.3$\pm$1.4 & 1.329$\pm$0.085 & 2.541$\pm$0.025 \\
		HD 216437b & 1353$\pm$5 & 0.32$\pm$0.03 & 63$\pm$6 & 48654$\pm$490 & 41.1$\pm$1.2 & 2.28$\pm$0.07 & 2.486$\pm$0.006 \\
		HD 219077b & 5471$\pm$52 & 0.769$\pm$0.002 & 56.0$\pm$0.5 & 46602$\pm$2001 & 181.5$\pm$0.9 & 10.549$\pm$0.076 & 6.22$\pm$0.04 \\
		GJ 832b$^{4}$ & 3657$\pm$104 & 0.08$\pm$0.06 & 246$\pm$22 & 46881$\pm$250 & 15.4$\pm$0.7 & 0.68$\pm$0.09 & 3.56$\pm$0.28 \\
		\hline
	\end{tabular}
		\\
		(1) \citet{30177}
		(2) \citet{huang18}
		(3) \citet{73526paper}
		(4) \citet{gj832}
\end{table*}

\subsection{Simulation Approach}

To determine the underlying occurrence rates of giant planets in our sample, we must determine the degree to which incompleteness afflicts our survey data.  As in numerous of our previous works \citep[e.g.][]{periodvalley, etaEarth, witt15}, we compute the detectabilities of planets by a simple injection-recovery technique.  In brief, we add the Keplerian signal of an artificial planet on a circular orbit to the existing RV data, then we attempt to recover that signal using a generalized Lomb-Scargle periodogram \citep{zk09}.  A planet is considered detected if it is recovered with FAP less than 1\% based on the FAP estimation in \citet{zk09}.  We considered planets with 100 trial orbital periods between 100 and 6000 days, bounding the region of ``cool Jupiters'' out to the time baseline of our AAPS data.  The mass of the simulated planet is increased until 99\% of the configurations (30 values of the orbital phase) at a given period are recovered successfully.  We repeated this process for recovery rates of 90\%, 70\%, 50\%, 30\%, and 10\%.  The result is an estimate of detectability (i.e. recovery rate) for a planet of a given mass and period.  

The occurrence rate of giant planets in our sample is then computed using the methods detailed in \citet{jupiters1} and \citet{jupiters2}.  In brief, for each detected planet, we estimate the probability of having detected that planet using the results of the injection/recovery simulations described above, summed over the entire sample.  This is accomplished by computing two quantities for each detected planet.  First, for the specific $P$ and mass of the detected planet in question, we calculate the completeness fraction $f_{c}(P,M)$ for the \textit{non-hosts} in the sample:

\begin{equation}
f_{c}(P,M) = \frac{1}{N_{stars}}\sum_{i=1}^{N} f_{R,i}(P,M),
\end{equation}

\noindent where $f_R(P,M)$ is the recovery rate as a function of mass at period $P$, and $N$ is the total number of stars not hosting a giant planet.  In this way, we account for the detectabilities for each star individually, at each of the 100 trial periods.  This yields a number between 0 and 1, representing the probability that a planet with a given ${P,M}$ would have been detected in the overall sample.  Second, we calculate the recovery rate $f_{R}(P_i,M_i)$ for each detected planet, at the $P$ and mass of that planet.  This is a number between 0 and 1, representing the probability of having detected that planet given the data for that star.  Usually (but not always) this number is unity; values of $f_{R}<1$ can be thought of as cases where we ``got lucky'' in detecting the planet.  These two quantities are then combined in Equation (2) to derive the number of expected detections given the data, and thence the number of ``missed'' planets:

\begin{equation}
N_{missed} = \sum_{i=1}^{N_{hosts}} 
\frac{1}{f_{R,i}(P_i,M_i)f_{c}(P_i,M_i)} - N_{hosts}
\end{equation}

\noindent where the symbols have the same meaning as given above.  The occurrence rate of planets in a sample is then first estimated as simply the number of detections divided by the total number of stars, using binomial statistics.  The completeness correction in Equation (2) is then used to boost the occurrence rates and their uncertainties by a factor $(N_{missed}+N_{detected})/N_{detected}$ to reflect the incomplete detection efficiency of our observational data.

\section{Results and Discussion}

We used the results of the injection-recovery simulations described above to address the two main questions posed in this paper: (1) What is the occurrence rate of giant planets as a function of orbital period, and (2) What is the frequency of \textit{pairs} of cool giants?  The first question seeks to probe the migration histories of these systems by searching for patterns in the orbital period distribution of cool giants in their final locations.  With the second experiment, we seek a preliminary understanding of how common scaled-down Solar system analogs might be (i.e. systems with two cool giants analogous to Jupiter and Saturn).  

\subsection{Period distribution of Jupiters hot and cool}

\begin{figure}
	\includegraphics[width=\columnwidth]{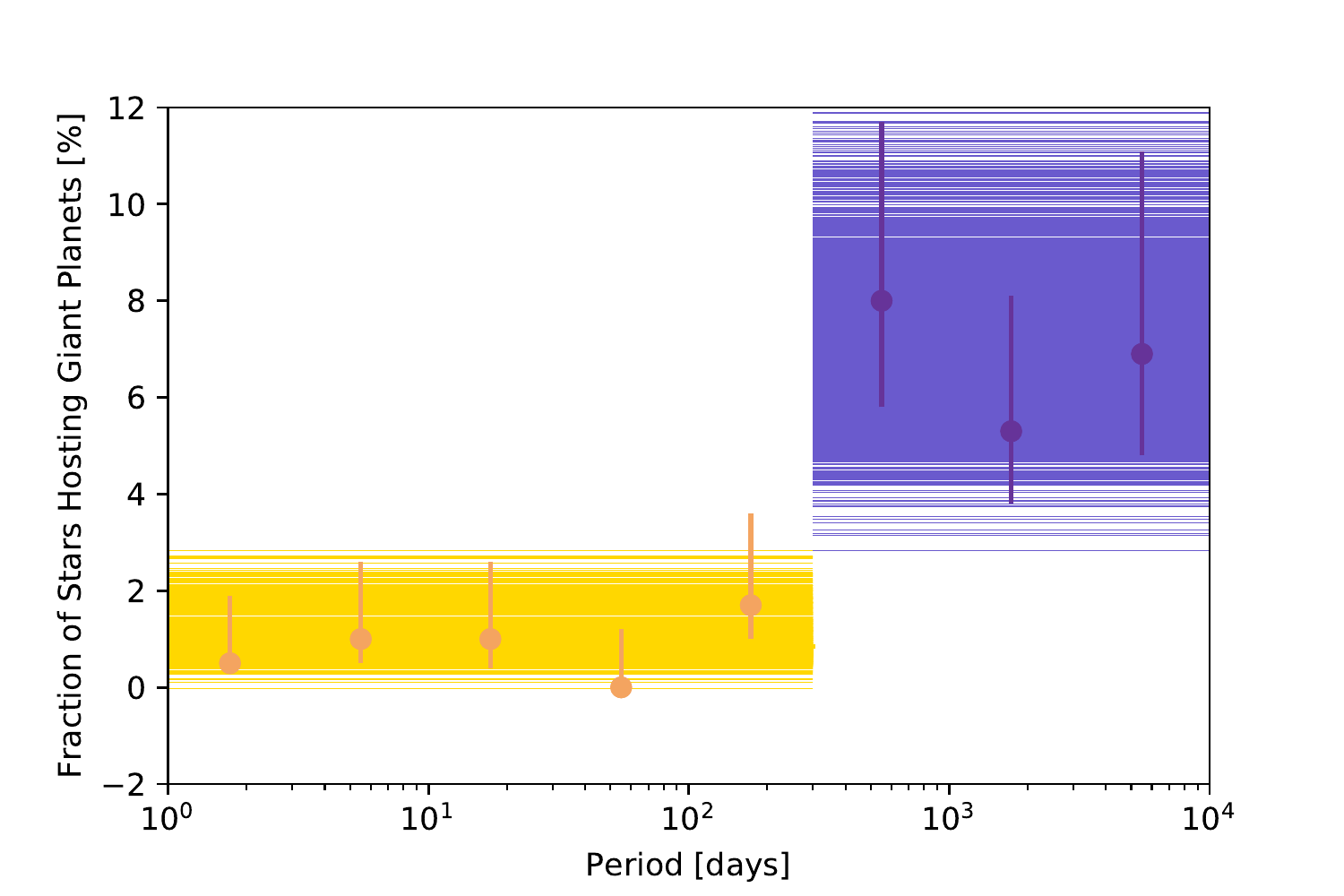}
    \caption{Frequency of giant planets as a function of orbital period.}
    \label{fig:frequency}
\end{figure}

To investigate the occurrence rate of giant planets as a function of orbital period, we divided the period range into evenly-sized bins in log space ($\Delta$log\,$P=0.5$).  We then computed the missed-planet corrections (Eq. 2) and occurrence rates in each bin based on the detections from our sample.  Table~\ref{tab:summary} summarises the results, and Figure~\ref{fig:frequency} shows the occurrence rates in their orbital period bins.

\begin{table}
	\centering
	\caption{Completeness corrections for the known giant planets in our dataset, as a function of their orbital period}
	\label{tab:summary}
	\begin{tabular}{lll} 
		\hline
		Planet & $f_R$ & $f_c$  \\
		\hline
		Period bin: 1-3 days -- 0.5$^{+1.4}_{-0.2}$\% & & \\
		HD 83443b & 1.00 & 1.0000 \\
		\hline
		Period bin: 3-10 days -- 1.0$^{+1.6}_{-0.5}$\% & & \\
		HD 179949b & 1.00 & 1.0000 \\
		HD 75289b & 1.00 & 1.0000 \\
		\hline
		Period bin: 10-30 days -- 1.0$^{+1.6}_{-0.5}$\% & & \\
		HD 108147b & 1.00 & 1.0000 \\
		HD 13445b & 1.00 & 1.0000 \\
	    \hline
		Period bin: 30-100 days -- 0.0$^{+1.2}_{-0.0}$\% & & \\ 
	    \hline
		Period bin: 100-300 days -- 1.7$^{+1.9}_{-0.7}$\% & & \\
		HD 208487b & 0.95 & 0.8052 \\
		HD 73526b & 1.00 & 0.9980 \\
		HD 134987b & 1.00 & 0.9780 \\
	    \hline
		Period bin: 300-1000 days -- 8.0$^{+3.7}_{-2.2}$\% & & \\
		HD 17051b & 0.96 & 0.9953 \\
		HD 160691e & 1.00 & 0.7250 \\
		HD 159868c & 1.00 & 0.8737 \\
		HD 142b & 1.00 & 0.9641 \\
		HD 38283b & 0.50 & 0.4448 \\
		HD 73526c & 1.00 & 0.9917 \\
		HD 154857b & 1.00 & 0.9875 \\
		HD 27442b & 1.00 & 0.9656 \\
		HD 20782b & 0.95 & 0.9568 \\
		HD 160691b & 1.00 & 0.9651 \\
		HD 23079b & 1.00 & 0.9859 \\
		HD 213240b & 1.00 & 0.9984 \\
	    \hline
		Period bin: 1000-3000 days -- 5.3$^{+2.8}_{-1.5}$\% & & \\
		HD 187085b & 1.00 & 0.7565 \\
		HD 2039b & 1.00 & 0.9964 \\
		HD 159868b & 1.00 & 0.9674 \\
		HD 23127 & 0.94 & 0.9254 \\
		HD 216435b & 1.00 & 0.9005 \\
		HD 216437b & 1.00 & 0.9622 \\
		HD 196050b & 1.00 & 0.9845 \\
		HD 39091b & 1.00 & 1.0000 \\
		HD 70642b & 1.00 & 0.9306 \\
		HD 30177b & 1.00 & 1.0000 \\
	    \hline
		Period bin: 3000-10000 days -- 6.9$^{+4.2}_{-2.1}$\% & & \\
        HD 154857c & 1.00 & 0.9467 \\
        GJ 832b & 1.00 & 0.4672 \\
        HD 114613b & 1.00 & 0.2092 \\
        HD 160691c & 1.00 & 0.8990 \\
        HD 134987c & 1.00 & 0.5072 \\
        HD 219077b & 1.00 & 0.9995 \\
        HD 142c & 1.00 & 0.9944 \\
        HD 30177c & 1.00 & 0.9990 \\
		\hline
	\end{tabular}
\end{table}

In previous studies, the existence of a ''period valley'' in the distribution of exoplanet orbits has become well established \citep[e.g.][]{jones03, udry03, periodvalley}. That ''valley'' is a range of orbital periods in which few planets are detected -- ranging between $\sim$ 30 and $\sim$ 100 days. The ''period valley'' can be clearly seen in our data, in the form of an empty bin for periods between 30 and 100 days.

For periods $P<400$ days, our results are consistent with the $\sim$4\% occurrence rate found by \citet{santerne16}.  The occurrence rates of giant planets plateaus at longer periods, but does not appear to fall off, in contrast to the findings of \citet{fernandes18}, who found evidence for a turnover in the occurrence rates at periods of 1000-2000 days.  However, our sample size is considerably smaller, such that within our uncertainties (Figure~\ref{fig:frequency}), we cannot exclude the fall off at the snow line postulated by \citet{fernandes18}.  For the ``Jupiter analogs'' at periods $P>3000$ days, our results remain consistent with the literature \citep[e.g.][]{cumming08, zech13, rowan16, jupiters2}.  

\citet{cumming08} and \citet{petigura18} found an increase in occurrence by a factor of $\sim$5 in the 100-300 day bin compared to shorter periods, but we find that bin enhanced by at most a factor of two, with large uncertainties.  This is almost certainly due to the small number of planets in our sample (\citet{petigura18} used the entire California Planet Search and {\textit{Kepler}} samples).  We note that the AAPS sample of long-period giant planets has the potential to be expanded with further observations \citep{legacy} from other precise RV facilities such as {\textsc{Minerva}}-Australis \citep{RobMin,addison19}.  Consistent with \citet{petigura18} and \citet{cumming08}, we do find a generally increasing giant-planet occurrence rate with orbital period.    

\subsection{Do cool Jupiters come in pairs?}

Of the 203 AAPS stars considered here, 25 (12.3\%) have at least one cool Jupiter, and 7 (3.4\%) have multiple such planets.  That is, if a star hosts one cool giant, it has a $\sim$25\% probability of hosting additional cool giants that could be detected using current methods.  We used the injection-recovery simulations as above to correct for incompleteness.  One subtle difference here is that, now that we are considering multiple-planet systems, the identity of the ``detected'' planet in Equation 2 becomes ambiguous.  We repeated the calculations considering both the first and second giant planets discovered.  One can imagine that, in considering the second detected planet, the recovery rate $f_{R}(P_i,M_i)$ for its period and mass would be lower than that for the first detected planet (i.e. the second planet is less obvious and hence harder to detect).  Indeed we find this to be the case: when considering the second detected planet, we arrive at 1.27 missed planets among the 7 multiples, whilst we only estimate 0.15 missed planets when considering the first planet discovered in the multiple systems.  When applying these correction factors to the binomial statistics as in the previous subsection, we arrive at statistically identical occurrence rates of cool-giant pairs regardless of which is detected first: 4.0$^{+2.7}_{-1.3}$\% (second planet) and 3.5$^{+2.3}_{-1.1}$\% (first planet).  

\citet{gould10} presented the first analysis of the occurrence rate of multiple giant planets beyond the snow line.  Based on microlensing results, including one detection of a Jupiter/Saturn analog pair \citep{gaudi08}, they estimated that about 1/6 (17\%) of stars host multiple cold giants akin to our Solar system.  Although that estimate is somewhat higher than that which we obtain in this work, it is worth remembering that detections using microlensing are sensitive to planets at very large orbital radii, whilst the data set presented herein is only sensitive to planets with orbital periods P$\ltsimeq$\,6000 days ($a\sim$7-8\,au).  As such, it seems reasonable to assume that as we become ever more sensitive to planets of lower mass and planets at larger orbital radii, the planetary abundances we determine will continue to climb.

In the regime of small samples and/or high incompleteness, the derived occurrence rates can be sensitive to the injection/recovery techniques.  Improved techniques and higher-quality data can result in increased planet occurrence rates (which, strictly speaking, ought not to happen if the incompleteness corrections are performed perfectly).  For example, the early \textit{Kepler} planet occurrence rate papers showed a turnover in the frequency of small planets, with a peak near $\sim$2.5\REarth\ and a drop for smaller planets \citep[e.g.][]{howard12,fressin13,petigura13}.  Subsequent analysis on the \textit{Kepler} Q1-Q16 data by \citet{burke15} showed instead a monotonically rising occurrence rate down to $\sim$1\REarth\ (cf. their Figure 8).  The latter work directly injected the simulated planets into the raw photometry and processed the data through the usual pipelines, to produce a more robust result.  Similarly, it is possible that the derived occurrence rates for long-period giant planets may increase in future as new data reduce the impact of incompleteness.  For long-period, low-amplitude planets, astrophysical and instrumental factors may also affect detectability.  Stellar magnetic activity cycles are a significant concern when considering RV signals of periods 10-20 years and amplitudes of only a few \ms\ \citep{johnson16,yee18}, since such signals can easily be produced both by activity cycles and by Saturn analogues (the RV signal of Saturn is 3\,\ms\ over 30 years).  Long-term RV surveys must also be careful in accounting for small ($\sim$1\,\ms) offsets caused by instrumental factors such as hardware upgrades or aging calibration lamps \citep{locurto15,talor19}. 

In the coming years, as surveys such as \textsc{Minerva}-Australis \citep{RobMin,addison19} take the mantle of continuing the work of the old radial velocity surveys (such as AAPS), we expect to be able to extend our study to longer orbital periods.  The improved sensitivity of those new instruments (which aim to achieve precisions of $\sim$ 1\,\ms\ for quiet stars), combined with the wealth of astrometric data that should become available from the \textit{Gaia} spacecraft, it seems likely that we will be able to extend our work to consider the abundance of less massive planets, as well as those on still longer-period orbits, working towards the detection of true analogues for the Solar system's ice-giants, Uranus and Neptune.

\section{Conclusions}

A key legacy of the 18-year Anglo-Australian Planet Search is the high efficiency with which that survey detected giant planets moving on long-period orbits.  Whilst other more recent RV surveys have achieved better measurement precision, and hence have discovered and characterised a great many low-mass super-Earths and sub-Neptunes \citep[e.g.][]{udry19, borsato19, feng19}, the venerable AAPS has greatly contributed to our understanding of cool Jupiters.  There is ultimately no substitute for time.  

In this work, we have examined the occurrence rates of such planets across a wide swathe of orbital period space.  Our main conclusions are:\\
1). We find that giant planets are about eight times more commonly found in orbits beyond about 1\,au than in closer-in orbits. \\
2). There are no significant differences in the occurrence rate of gas giants beyond $\sim$1\,au.

Although it is still unclear why gas giants with orbital distance within $\sim$1\,au exist at all, the preservation of this sharp transition implies that planet-planet scattering (due to dynamical instabilities) has not greatly altered the kinematic structure of planetary architecture.  

The transition semi-major axis may correspond to the snow line, which implies the inner boundary of the birth domain.  That we see a transition near 1\,au is somewhat unexpected based on the classic definition of snow line \citep{Hayashi1981}, which is about 2.7\,au away from the Sun. \citet{Sasselov2000}, however, found that snowlines could be as close as 1\,au to the central stars.  Even so, whether the closer region is a forbidden zone for gas giant formation is still under severe debate \citep{Batygin2016}. Our result implies that gas giants are less likely to be formed in the inner region of a system, but it is still possible.  However, the big conflict between the in-situ formation model and current observational results is: Hot Jupiters appear to be as common as warm Jupiters.  Hot super-Earths, however, are much less common than cold super-Earths (See \citealt{Dawson2018}).  

The transition semi-major axis may also be the location where photoevaporation of the disk gas occurs.  It may lead to divergence in the migration outcome (i.e., planets inside it migrate inwards and planets outside it migrate outwards).  But the process corresponding to that timescale of migration is similar to (or slightly longer than) the timescale of disk depletion.  There are some uncertainties, however, on these two migration time scales \citep{Ida2004}. 

In the future, as the true distribution and occurrence of planets at larger orbital radii becomes clearer, data such as ours will doubtless prove vital in constraining different models of exoplanet formation.  Rather than simply explaining a given population of exoplanets (such as the hot Jupiters), such results will allow theorists to build models of planet formation that cover a wide variety of initial conditions - from stellar mass and disk mass to the metallicity of the system as it forms. 

Finally, we note that while the focus of interest in exoplanetary science has been on the revolution wrought by the space-based transit missions of \textit{Kepler} and \textit{TESS} \citep{kep1,tess1}, and further exciting insights will undoubtedly be gained from future missions, there remains a need to continue the legacy RV work highlighted here.  For it is only with such continued, daresay a multi-generational effort, that we will come to understand the \textit{complete} architectures of planetary systems, probing out to Saturn analogues and beyond.  And at the end of all our exploring, we will come to a better understanding of our own Solar system in the Galactic context.    


\section*{Acknowledgements}

We acknowledge the traditional owners of the land on which the AAT stands, the Gamilaraay people, and pay our respects to elders past and present.  S.W. thanks the Heising-Simons Foundation for their generous support as a 51 Pegasi b fellow.  We thank Douglas Lin for helpful discussions.  This research has made use of the NASA Exoplanet Archive, which is operated by the California Institute of Technology, under contract with the National Aeronautics and Space Administration under the Exoplanet Exploration Program.  This research has made use of NASA's Astrophysics Data System.  This material is based upon work supported by the National Science Foundation under Grant No. 1559487.








\appendix

\section{Some extra material}

If you want to present additional material which would interrupt the flow of the main paper,
it can be placed in an Appendix which appears after the list of references.


\bsp	
\label{lastpage}
\end{document}